\documentclass[draft]{agujournal}
\draftfalse

\journalname{Journal of Geophysical Research}

\begin{document}

%
%


\title{Separator reconnection at Earth's dayside magnetopause:  MMS observations compared to global simulations}

%
%




\authors{Natalia Buzulukova \affil{1,2}, John. C. Dorelli \affil{1}, Alex Glocer \affil{1}}

 \affiliation{1}{NASA Goddard Space Flight Center, Greenbelt, Maryland, USA}
\affiliation{2}{Astronomy Department, University of Maryland College Park, College Park, Maryland, USA}





\correspondingauthor{Natalia Buzulukova}{natalia.y.buzulukova@nasa.gov}




\begin{keypoints}
\item  MMS observations of a southward IMF EDR encounter are consistent with the magnetic separator location as predicted by global MHD simulation.
\item The model predicts a complex magnetic topology in the vicinity of the MMS, with multiple separators and flux ropes passing by the MMS s/c.
\item  The simulation shows that the existence of IMF Bx results in a duskward shift of the location of the topological separator.
\end{keypoints}

%
%


\begin{abstract}
We compare a global high resolution resistive magnetohydrodynamics (MHD) simulation of Earth's magnetosphere with observations from the Magnetospheric Multiscale (MMS) constellation for a southward IMF magnetopause crossing during October 16, 2015 that was previously identified as an electron diffusion region (EDR) event.
The simulation predicts a complex time-dependent magnetic topology consisting of multiple separators and flux ropes.
Despite the topological complexity, the predicted distance between MMS and the primary separator is less than 0.5 Earth radii.
These results suggest that global magnetic topology, rather than local magnetic geometry alone, determines the location of the electron diffusion region at the dayside magnetopause.
\end{abstract}

%
%

%


%
%
%
%

\section{Introduction}

Magnetic reconnection is the primary mode by which the solar wind couples to Earth's magnetosphere, driving the Dungey convection cycle \citet{dungeya,dungeyb} that powers magnetic storms, the aurora and global energy circulation in Earth's magnetosphere. 
Our understanding of the local properties of magnetic reconnection has increased steadily over the last two decades.
The importance of the Hall effect in decoupling the reconnection rate from the plasma resistivity in the low resistivity limit was recognized early by \citet{biskampc} and \citet{shayb} and culminated in the GEM reconnection challenge \citep{birn2001}.
The role of Speiser orbits \citep{speiser1965} and the resulting non-gyrotropic electron velocity distributions \citep{vasyliunas1975} in supporting the reconnection electric field in the absence of collisions was confirmed by two-dimensional PIC simulations and further developed by  \citet{hessea}.
Later, larger scale three-dimensional PIC simulations showed how collisionless electron tearing and associated plasma turbulence significantly alters the structure of the reconnection layer \citep{daughtond}.
The effects of current sheet asymmetry  \citep{CassakAssymetric} and the guide field \citep{rogers2003}  were explored for dayside magnetopause applications.

Three-dimensional reconnection theory, on the other hand, is in an earlier stage of development, and the very definition of three-dimensional reconnection is still a subject of vigorous debate  \citep{greenea,schindlera,hessed,laua,boozerb,dorellif,dorelli2008}.

Nearly everyone agrees that two-dimensional reconnection occurs at locations where the magnetic field, projected into the plane perpendicular to the ignorable coordinate, vanishes (note that even in cases where there is no true magnetic null, reconnection occurs along a "guide field" which is a projected X type null analogous to a closed field line in a toroidal system).
From this perspective, it is natural for spacecraft observations to focus on local geometrical properties of the magnetic field (e.g., magnetic curvature forces which accelerate the plasma) and plasma (e.g., signatures of unmagnetized electrons).

Three-dimensional reconnection is more difficult to identify because changes in magnetic topology (e.g., a field line changing from "open" to "closed") are not always reflected in the local magnetic geometry at some point; similarly, changes in magnetic geometry (e.g., local twisting of a magnetic flux bundle) need not result in a change in magnetic field topology.
The implication is that spacecraft observations that focus on locally observable changes in magnetic field geometry (e.g., by measuring plasma acceleration or locally reconstructing the magnetic field and pressure gradients) may miss the topological structures that actually control the global reconnection rate.

NASA's Magnetospheric Multiscale (MMS) mission \citep{burch2016}  is uniquely suited to addressing the problem of three-dimensional reconnection.
First, the mission was designed to directly measure the properties of the electron diffusion region -- specifically, the electron gyro-viscous signatures that  expected to occur only near the topological separator -- independently of observations of the local magnetic geometry.
Second, MMS has now crossed the dayside magnetopause thousands of times, making possible for the first time a study of the global structure of the reconnection dissipation region.
Such global studies of collisionless reconnection are still out of reach of the most powerful supercomputers, though we will see below that much can still be learned from global magnetohydrodynamics (MHD) simulations.

MMS has already observed with unprecedented spatial and temporal resolution many of the kinetic scale features predicted by local kinetic simulations:  the decoupling of electron and ion bulk velocities in the ion diffusion region (e.g., \citet{lavraud2016}); evidence of filamentary current structures \citep{phan2016} ; ion-scale magnetic flux ropes \citep{eastwood2016} ;
and crescent shaped non-gyrotropic distribution functions near the reconnection stagnation point \citep{BurchScienceEDR,chen2016}.

The observations presented by \citet{BurchScienceEDR} are particularly compelling since they are the first direct observations of plasma dissipation at an electron-scale current sheet near a reconnection site at the magnetopause.
On October 16, 2015, MMS observed a clear ${\bf J} \cdot ({\bf E} + {\bf V}_e \times {\bf B/c})$ signal (where {\bf J} is the current density, {\bf E} is the electric field, ${\bf V}_e$ is the electron bulk velocity, {\bf B} is the magnetic field vector, and c is the speed of light) at the current sheet, and the dissipation appeared to involve mostly perpendicular current carried by crescent shaped electron velocity distributions.
Many of the plasma and field structures seen in the data were consistent with those predicted by two-dimensional particle-in-cell (PIC) simulations.

MMS has now observed a handful of similar electron dissipation region encounters, so that it is now possible to directly compare locally observed signatures of topological dissipation with those predicted by global simulations.
In this paper, we describe a first step in this direction, comparing the locations of the topological separators predicted by a high resolution global magnetohydrodynamics (MHD) simulation with the location the October 16, 2015 MMS electron dissipation region encounter. 

\section{Global simulations}

Our global magnetosphere simulations are performed using the BATS-R-US global MHD code developed at the University of Michigan  \citep{powell1999,toth2005,toth2012}.
For this study, we make use the isotropic resistive MHD version of BATS-R-US with uniform resistivity.
The choice of resistive MHD is motivated by the need to control the numerical resistivity and achieve as high a Lundquist number as possible while still resolving the resistive dissipation region.
For the simulation described here, we use the value of resistivity $\eta =  2.125 \times 10^9 m^2$/sec and grid resolution at the dayside magnetopause region $\Delta$R=1/16 R$_E$.  
The original BATS-R-S grid was changed to adopt fine grid resolution in a region containing the dayside magnetopause. 
Our choice of parameters allows us to produce relatively thin (on MHD scale) current sheets at the magnetopause ($\sim 0.5 R_E$) , flux ropes and Flux Transfer Events (FTEs) under southward Interplanetary Magnetic Field  (IMF) (e.g., see \citet{Glocer2016}). 
The grid resolution is chosen to be 1/8 R$_E$ in the box XYZ= -10/+10 R$_E$ in order to properly resolve the near-Earth region.  
The simulation box is defined by X=[-224,32]; Y=[-128,128]; Z=[-128,128] $R_E$. 
The inner boundary is set up at r=2.5 R$_E$.  
GSM coordinate system is used, and the total number of cells is $\sim 3 \times 10^7$.

For global MHD simulations of Earth's magnetosphere, the MHD module is normally coupled with the ionospheric electrodynamics module. 
Since we are simulating a real event, we choose the ionospheric conductivity model to be a sum of a UV component controlled by the f10.7 index and an event-dependent conductivity from the empirical model of \citet{ahn1983} and \citet{ridley2004}. 
Earth's dipole tilt and rotation are also included.


We perform the run for the October 16, 2015 event for the time interval 1100 UT - 1315 UT in order to  see the dynamics around $\sim$1307 UT where MMS s/c detected EDR. 
IMF Bz during the EDR encounter was slightly negative with Bz $\sim$ - 3 nT.  
The IMF By, solar wind density and temperature were taken from the OMNI database with 1 min resolution and used as a boundary condition for the model at X=32 R$_E$. 
IMF Bx was kept constant at Bx=2 nT (close to IMF Bx near 1307 UT) in order to eliminate any possible numerical effects from magnetic field divergence error. 
However, as the results show, both Bx and By components are crucial since they determine (through the draping of the IMF) the distribution of the current density -- and, hence, the location of the topological separator -- on the magnetopause. 

We have developed a new, efficient algorithm for tracking magnetic separator in three-dimensional global magnetosphere simulations \citep{Glocer2016}.
The software is called RECON-X and is now available to the community at the Community Coordinated Modeling Center (CCMC) as a post-processing option for global MHD simulations.
The algorithm is an efficient "divide and conquer" algorithm that avoids the need to exhaustively sample the simulation domain to computer global magnetic field topology.
Instead of brute force computation of field lines sampled on the plane (impractical, particularly for very large data sets like those produced by BATS-R-US), we subdivide a given plane into large squares and sample the field topologies only on the edges of the squares.
Four magnetic field topologies are defined:  1) solar wind, 2) open and intersecting the northern ionosphere, 3) open and intersecting the southern ionosphere, 4) closed and intersecting both ionospheres.
Squares whose edges have all four topologies are subdivided, and the process is repeated until some accuracy criterion is achieved.
Intuitively this algorithm  could be envisioned as seeking the points of line, where two 3D objects intersect each other: the object that contains closed field lines and the object that contains open field lines. These objects  are also referenced as surface of 'last closed field lines' add the surface of 'first open field lines'.
For further details of algorithm and a discussion of their optimization please see \citet{Glocer2016}.   
Since the procedure is computationally expensive, we find the separator points in the volume  XYZ= -14/+14 R$_E$ , separately for cuts in Z and Y planes, with dl=0.25 R$_E$ spacing between planes.  
    
\begin{figure}[h]
\centering
 \includegraphics[width=45pc]{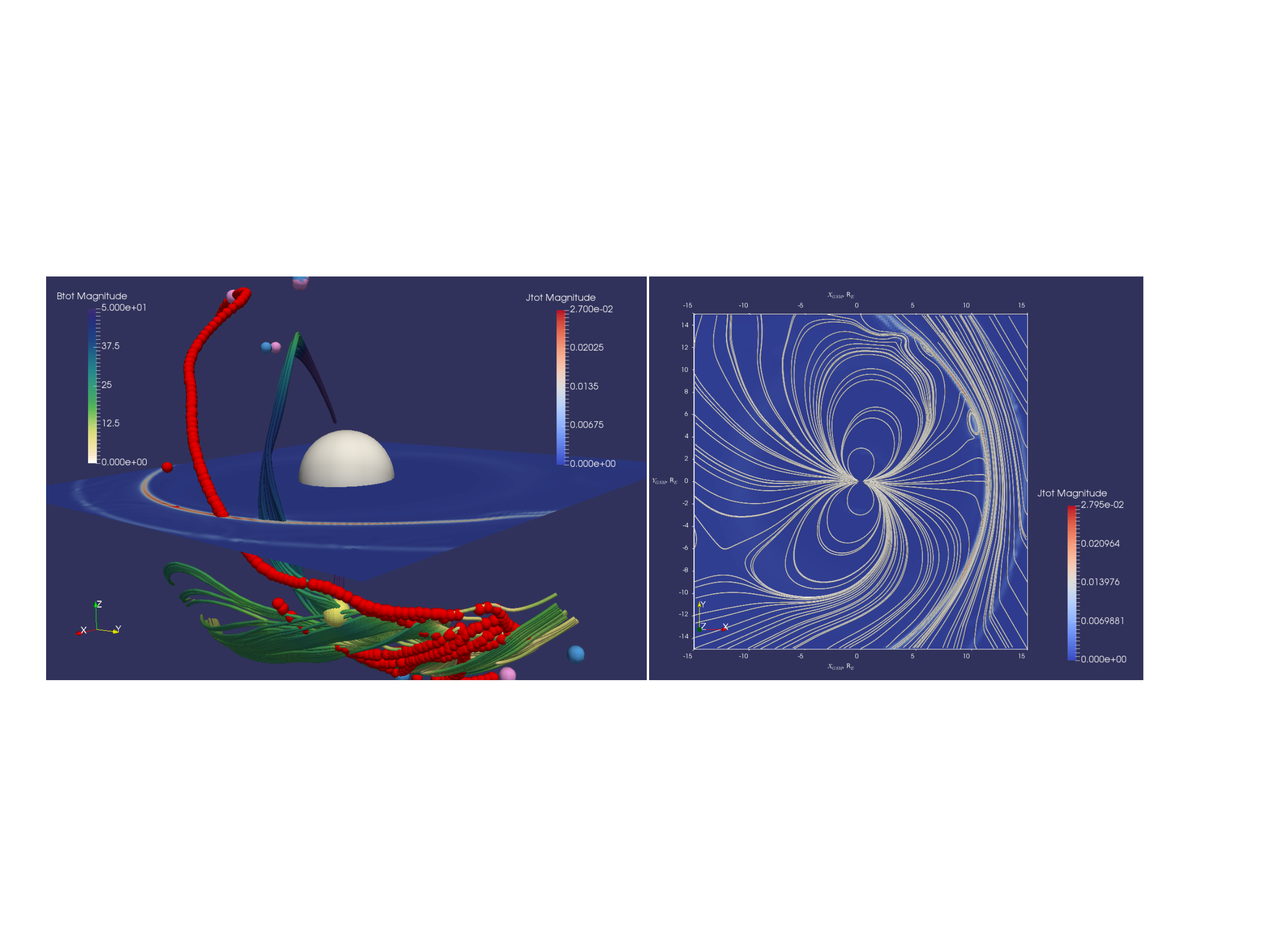}
%

\caption{Simulation results for October 16, 2015, 0100 UT.  The left panel: 3D view of the  separators topology and magnetic nulls. 
The MMS tetrahedron location is shown by yellow sphere; the separators are shown  by red spheres. 
Positive nulls are pink spheres and negative nulls are blue spheres.  The cut (Z=0 GSM) shows total current density.  
A magnetic flux rope is evident around the satellite location. 
The central sphere has radius r=3 R$_E$. 
The right panel shows the solar wind magnetic flux pile-up effect.  The sun is on the right.
The 2D projections of magnetic lines in the plane Z=0 GSM shows the magnetic draping effect near the magnetopause. 
This effect determines the distribution of the total current density at the magnetopause, causing a skewing of maximal current density and separator to the dusk side.  
The field line projections are plotted over 2D map of current density.}
 \label{fig1}
\end{figure}

The results of the MHD simulation and post-processing with RECON-X for 0100 UT are shown in Figure~\ref{fig1}.  
The left panel shows the global 3D view of the separator lines (red spheres) together with the magnetic nulls (pink and blue spheres). 
The cut plane Z=0 GSM is drawn to show the distribution of total current density (jtot) at the magnetopause.  
There is a set of magnetic field lines to highlight the FTE passing through the MMS location around that time. 
The magnetic filed lines are colored by magnetic field strength.  
A central gray sphere with r=3R$_E$ is included to visualize the near-Earth region.  
The  yellow sphere shows the location of the MMS tetrahedron.  
The right panel shows the effect of magnetic field line draping around the Earth and the creation of a magnetic flux pile-up region.  
The draping/pile up effect changes the distribution of current density along the magnetopause by adding gradients to the magneto-sheath  magnetic field. 
For the given configuration of the IMF Bx and By, the draping/pile up intensifies the current density in the dusk sector and shifts the location of current sheet maximum and, as we will see, the topological separator.

Interestingly, the simulation shows a very complex magnetic topology during the interval when MMS crossed the magnetopause.
The existence of multiple separators is a consequence of the formation of flux ropes and FTEs, and this is consistent with the previous results of \citet{dorelli2008} and \citet{Glocer2016}.

Simulation results for 0107 UT are shown in the Figure 2.  
At this time, the MMS tetrahedron encountered the EDR \citet{BurchScienceEDR}. 
Similar to the Figure 1, there are also multiple separators indicating a complex magnetic field topology.  
Here MMS was located very close to the separator points (< 0.5 R$_E$).  By this time, the previous flux rope has already passed by the tetrahedron, but flux ropes continue to form at roughly 5-10 minute intervals throughout the simulation.

\begin{figure}[h]
\centering
 \includegraphics[width=45pc]{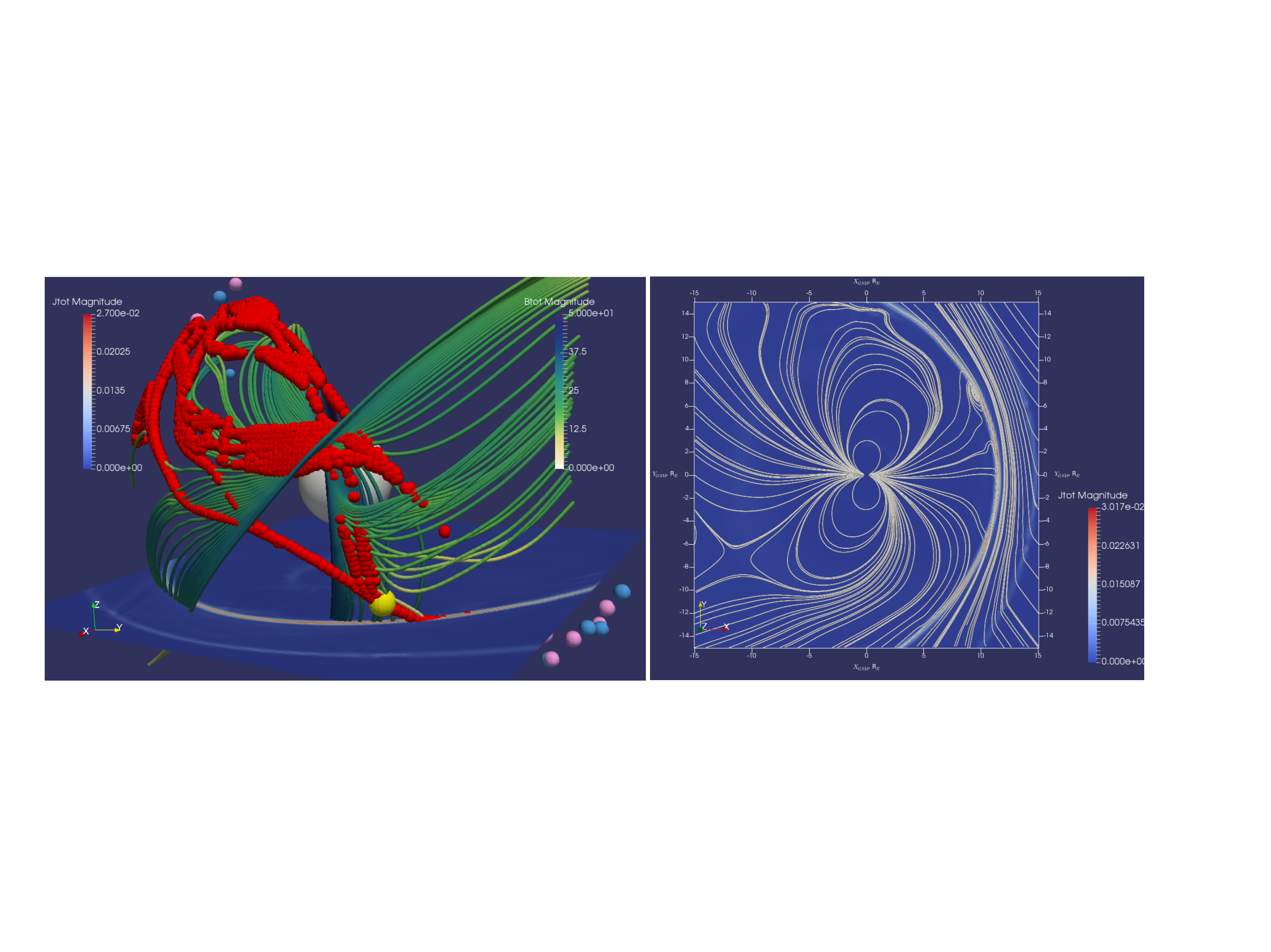}
%

\caption{Simulation results for October 16, 2015, 0107 UT.  
The format is the same as Figure 1. 
The cut plane in the left panel is drawn for Z=-5 R$_E$ GSM. 
The cut plane for the right panel is drawn for Z=0 (GSM). 
The MMS tetrahedron is located in the vicinity of the  separator (< 0.5$R_E$). 
Note that at the same time (0107 UT) MMS encountered the EDR.  
In comparison with the results for 0100 UT, there is no flux rope structure around the s/c at this time.}
 \label{fig2}
\end{figure}

As mentioned above, the draping/pile up effect should change the distribution of current density on the magnetopause.  
Figure 3 shows the location of maximal current density, plotted as a function of MLT, defined in two cut planes Z=0 and Z=-4.8 $R_E$ GSM (drawn through MMS location).  The formation of flux ropes  distorts the background current density, causing temporary drops at particular local times (this effect was also observed by  \citet{Glocer2016}).   
Despite this current disruption effect of the simulated flux ropes, the profile of current density along MLT for Z=0 nevertheless shows a clear asymmetry in the background current density. 
The region with enhanced current density is not symmetric around the noon -- it is shifted to the dusk sector (for Z=0).  
For 1258 UT, Z=0 the  left part of the plot (dawn sector) is undisturbed and  it is different from the right undisturbed part of the profile at 0107 UT, Z=0.    
Another interesting observation is that all major disturbances in  the current sheet for the plane Z=0 are found in post-noon and dusk sectors. 
This asymmetry is explained by two factors: i) there are preferable conditions for instability in the dusk sector since current sheet is shifted because of draping/pile up effect; ii) the flux ropes are moving predominantly toward the evening sector in the background flow around the magnetopause.

The results for the cut plane  Z=-4.8  show the movement of the flux rope through the s/c location at $\sim$ 0100 UT and existence of two flux ropes around the s/c  at  $\sim$ 0102 UT.  
Around 0107 UT, when MMS detected the EDR crossing, the following is observed:  (i)  MMS is close to the separator;  (ii)  there is single  separator  at  MMS location at that time; (iii) current density has a maximum value along the magnetopause near the MMS and separator location. 
We emphasize this result: in the absence of multiple separators, the current density  maximizes near the separator location for a given  cut plane.

\begin{figure}[h]
\centering
 \includegraphics[width=30pc]{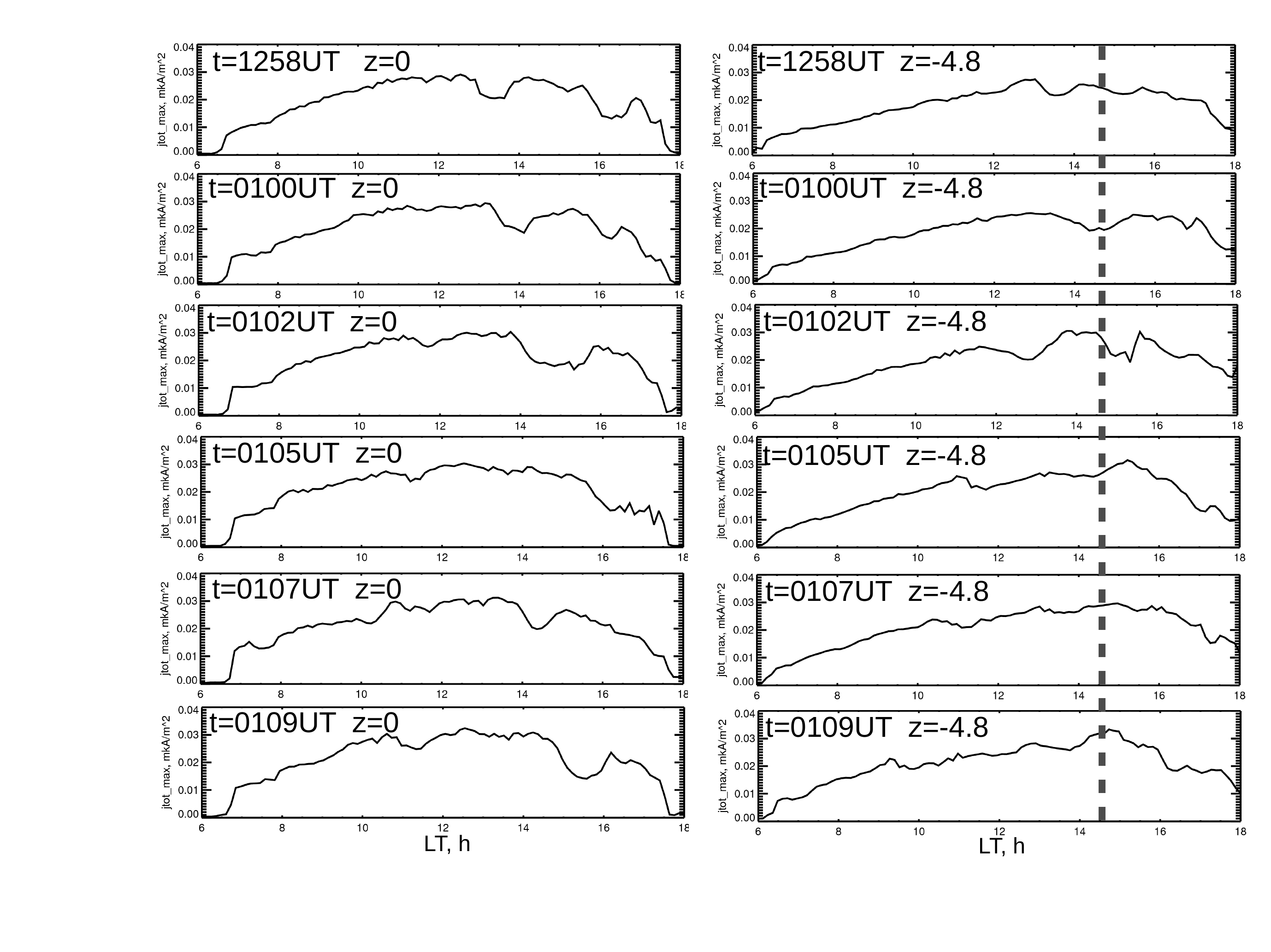}
%

\caption{The location of maximal current density as a function of GSM local time for the two cut plans, Z=0  and Z=-4.8 R$_E$ GSM, drawn through MMS location. 
The plots are shown for 6 time steps. 
The vertical line at LT$\sim$14.7h in the right panel denotes MMS location.}
 \label{fig3}
\end{figure}

\section {Putting the MMS October 16, 2015 EDR encounter into global context}
\subsection{Anti-parallel or guide field reconnection?}
\citet{BurchScienceEDR} show that the magnetic field magnitude Btot was very close to 0 near the EDR, < 1-2 nT around 0107 UT, with  Btot $\sim$ 25nT just 30 sec before EDR crossing and Btot $\sim$20 nT 30 sec after EDR crossing. 
Correspondingly, the PIC simulations presented in  \citet{BurchScienceEDR} assumed an anti-parallel current sheet geometry. 
However, since the IMF at that time had substantial Bx$\sim$2nT and By$\sim$-2 nT  components (with Bz$\sim -3$ nT),  it is important to understand why anti-parallel geometry was observed for the October 16 EDR crossing.
The simulation results presented above may provide an explanation. According to the model results, during  interval 1255 - 1315 UT the MMS tetrahedron was located in the highly dynamic region with very complex field topology near the primary separator.  
In resistive MHD, the magnetopause acts as a thin resistive layer with  enhanced pressure (due to Ohmic heating) and hence depressed magnetic field.  
Example magnetic field profiles for 0107 UT and radial slice at Z=-4.8 $R_E$ and LT=14.7h are shown in Figure 4. 
The simulated magnitude of Btot in the middle of the current sheet is close to $\sim$ 15 nT, i.e. comparable with MMS observations outside the EDR. 
\begin{figure}[h]
\centering
 \includegraphics[width=30pc]{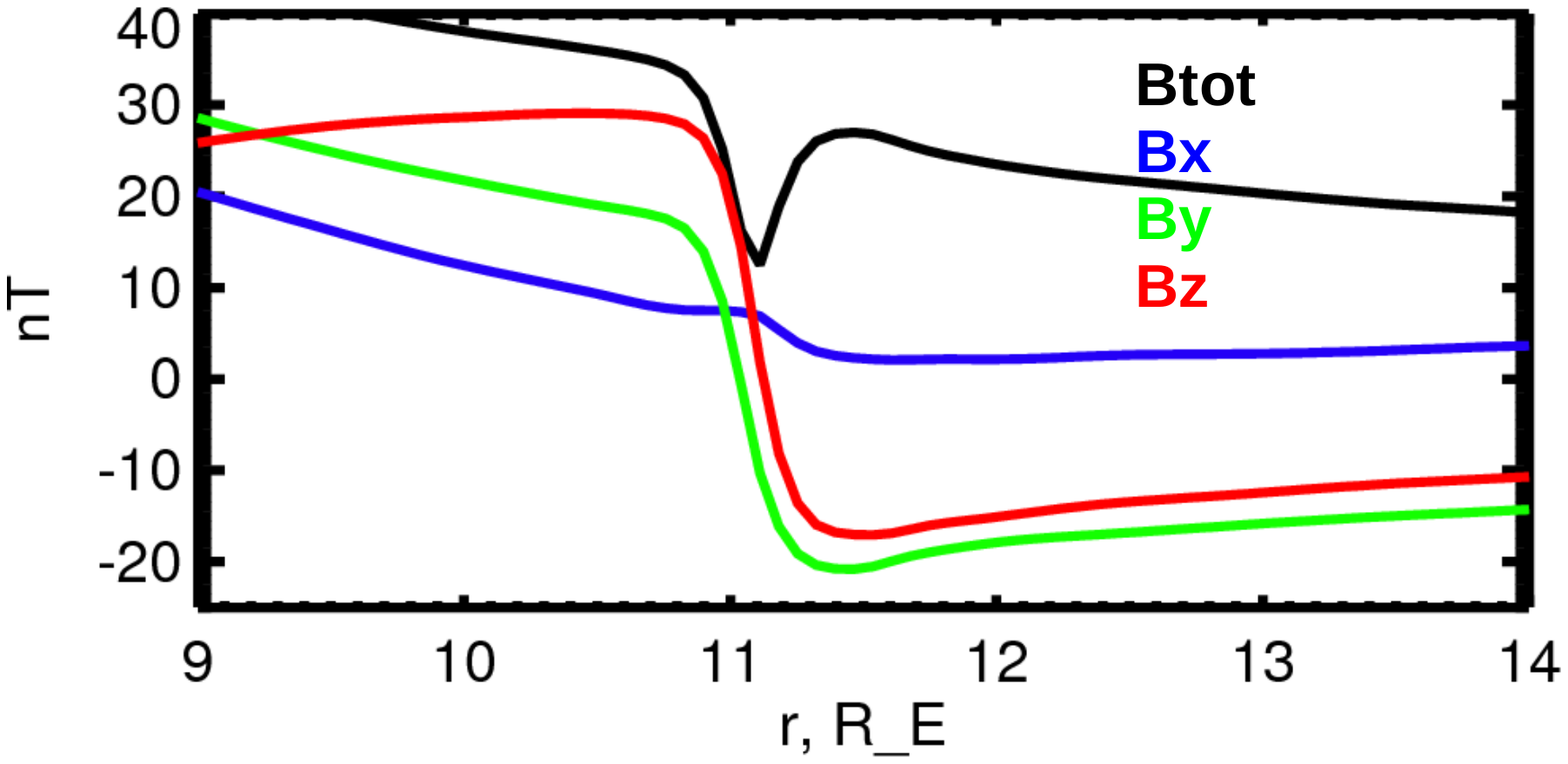}
%

\caption{Profiles of total magnetic field (black curve), Bx component of magnetic field (blue curve), By component (green curve) and Bz component (red curve) for t=0107 UT, Z=-4.8 R$_E$, LT=14.7 h cut. 
The magnetic field in the center of the current sheet  goes down to $\sim$ 15 nT.}
 \label{fig4}
\end{figure}

The simulation also shows a varying guide field along the separators with separators connecting magnetic nulls. 
However in the simulations, the closest magnetic nulls in the southern hemisphere are relatively far from the MMS location, at a distance of  a few $R_E$.     
At the same time,  the simulations show the formation of multiple flux ropes starting around a primary separator and moving predominantly in the  direction of the dusk sector, passing the location of the MMS tetrahedron. 
The flux ropes have a complex structure even in the MHD simulation,  without kinetic effects. 
We argue in the reality the kinetic effects could cause more complex magnetic field topology in  flux ropes,  bigger  magnetic field depression in thin kinetic scale current layer,  and/or  re-location of  MHD-derived magnetic nulls. 
All of these factors may contribute to the observed depression of Btot down to  $\sim$1-2 nT.  
PIC simulations also demonstrate that only a relatively small value of the guide field needs to be added, $\sim 0.1  B_0$  where $B_0$ is asymptotic magnetic field, to magnetize electrons around the EDR and change the reconnection regime from anti-parallel to guide field \citep{Swisdak2005}.

\subsection{Location of separators and IMF draping effect}

A common feature for the two time steps, 0100 UT and 0107 UT is the tree-like structure of separators. 
There is a  'trunk' and at some point  multiple `branches' of separator lines deviating from the `trunk'.  
We will call the `trunk' separator the `main',  or `primary' separator, and    
`branches' as `secondary' separators.  
The existence of a primary separator is routinely reproduced in global MHD models \citep{Dorelli2004,Dorelli2007, Hu2009,Komar2013,Hoilijoki2014,Komar2015}.  
We suggest secondary separators start to form near the primary separator as a result of instability.  
There is  continuous discussion how to  predict the location of the primary separator points (see e.g.  \citet{Komar2015} for the overview of the models). 
Since secondary separators are formed as a result of instability of thin current sheet, it is hard to expect their exact location from any model.  
For the primary separator points, we note there is only one model that directly takes into account the IMF draping effect and pile-up effect, namely the model of \citet{Alexeev1998}.
In this model, the reconnection is predicted to take place where the current density magnitude Jtot has a maximum.  
This is consistent with the results for 0107 UT where MMS location coincides with location of separator and also maximal current density (along cutting plane Z=-4.8 $R_E$).  
There are a number of conclusions that follow from this hypothesis and could be tested with future runs:

(i) since the reconnection rate is faster for southward conditions,  the pile-up effect and draping should be smaller than  for northward IMF Bz;

(ii) the draping and pile-up effects will be sensitive to the resolution: the better the grid resolution, the more pronounced the effect will be;

(iii) since the draping and pile up effects are sensitive to the orientation of IMF Bx and By,  both components must be taken into account to predict the location of the magnetic separator.

\section {Conclusions and Discussion}

In this paper, we have addressed the question of whether the topological separators predicted by global resistive MHD simulations are consistent with local EDR signatures observed by MMS.
Our conclusions are summarized as follows:

\begin{enumerate}
\item The location of MMS at the time of the October 16, 2015 EDR encounter reported by \citet{BurchScienceEDR} within about $0.5 \, R_E$ of the primary topological separator as predicted by the global MHD simulation.
\item The simulation shows that the existence of IMF $B_X$ results in a dawnward or duskward (depending on the sign of $B_X$) shift of the location of the topological separator, consistent with the MMS EDR encounter of October 16, 2015.
\item The simulation predicts that the magnetic field topology in the vicinity of MMS is complex, with multiple flux ropes, separators and magnetic nulls forming and passing over the MMS tetrahedron during the interval around the EDR encounter.
\end{enumerate}

These initial results suggest that global resistive MHD simulations can accurately predict the locations of electron dissipation regions at the dayside magnetopause.
A question immediately arises:  Why should we expect a resistive MHD simulation, containing none of the kinetic scale physics that controls the reconnection process, to make an accurate prediction of the location of the EDR?
We suggest that the answer lies in the association of the topological separator with the location of maximal current density on the magnetopause, a feature shared by both the global simulation as well as the MMS EDR events.
Indeed, one of the criteria used to identify MMS EDR candidates is local current density.
We further suggest that the location of the current density maximum on the magnetopause surface is determined by global draping and pileup of the IMF upstream of the magnetopause, which is an ideal MHD effect that should be insensitive to the details of the reconnection process.
This is not to say that kinetic processes do not impact the distribution (e.g., the EDR aspect ratio) and intensity of current density on the magnetopause; however, the location of the current density maximum may be insensitive to such details.

This above discussion raises another interesting avenue of future work.
One of the properties of the magnetopause current sheet that we expect to be sensitive to the collisionless reconnection physics is the aspect ratio of the EDR.
While local PIC simulations have suggested that the aspect ratio is always of order $\sim$10, it is very difficult to test how the aspect ratio scales up to very large systems like Earth's magnetopause (which is thousands of ion inertial lengths in size).  
If global MHD simulations can consistently predict the location of the EDR, then statistical analysis of the thousands of MMS magnetopause crossings may for the first time provide an answer to this question by allowing a study of the variation of locally observed EDR signatures as a function of distance from the predicted topological separator.

\acknowledgments
Natalia Buzulukova, John C. Dorelli and Alex Glocer acknowledge support from the following sources:  NASA  LWS Grant WBS 936723.02.01.09.47 and NASA HSR Grant  WBS 791926.02.01.07.22.






%
%
%
%
%
%
%
%
%
%

\bibliography{master.bib}





\listofchanges

\end{document}